# Ultra low-cost fabrication of homogeneous alginate hydrogel microspheres in symmetry designed microfluidic device

Qing Qin, Yu Zhang, Yubei Wei, Jinnuo, Lv, Meiling Tian, Yuanyuan Sun, Xingjian, Huang, Jianglin Li, Yifeng,Su, Xiaoliang Xiang, Xing Hu*, Zhizhi Zhou*

Key Laboratory of Research and Utilization of Ethnomedicinal Plant Resource of Hunan Province, Hunan Provincial Higher Education Key Laboratory of Intensive Processing Research on Mountain Ecological Food, College of Biological and Food Engineering, Huaihua University, Huaihua 418008, China

*Corresponding should be addressed to Xing Hu: huxing98@126.com; Zhizhi Zhou: dzhizhou@cqu.edu.cn;

**In this study, we present a two-stage method for fabricating monodisperse alginate hydrogel microspheres using a symmetrically designed flow-focusing microfluidic device. One of the flow-focusing junctions generates alginate hydrogel droplets without the addition of surfactants, while the other junction introduces corn oil with acetic acid, which facilitates the solidification of the homogeneous alginate hydrogel droplets and prevents coalescence. These hydrogel microspheres can be easily separated from the oil phase using an oscillation state, eliminating the need for a demulsifier. This microfluidic system for hydrogel microsphere formation is notable for its simplicity, ease of fabrication, and user-friendliness.**

Alginate hydrogel microbeads have been utilized in a diverse array of applications, ranging from academic research to industrial production[1, 2]. Traditional methods for the mass production of alginate microcapsules typically involve mechanical stirring or mixing through external gelation.[3-5]. However, these techniques often result in varied morphologies and dimensions of the microcapsules, complicating their applicability across different fields. In contrast, microfluidic techniques offer an alternative approach for producing monodisperse droplets with precisely controlled morphologies and sizes[6-9], thereby addressing the technological challenges associated with the aforementioned applications. The chaotic mixing of alginate solution and calcium ions at the T-junction site in microfluidic devices can lead to clogging due to the rapid cross-linking of microgels in narrow channels.[7] To circumvent this issue, calcium carbonate ($CaCO_3$) nanoparticles have been employed to produce microgels with a uniform size. Under acidic conditions, the water-insoluble $Ca^{2+}$ can be released into the alginate solution after emulsification[6]. However, the heterogeneous distribution challenging of large $CaCO_3$ particles leads to nonuniform cross-linking within the resulting microbeads. To address this issue, Weitz's group demonstrated the introduction of a water-soluble calcium-ethylenediaminetetraacetic acid (calcium-EDTA) complex to regulate the delivery of calcium ions, thereby achieving excellent homogeneity in microgel formation without inducing nonuniform gelation[8, 10]. The fabrication of cross-linked microgels involves multiple stages, which require labor intensive. Lin's group developed an integrated microfluidic system comprising two separate chips that produce high-quality hydrogel microbeads utilizing a calcium-EDTA complex as a precursor. This system facilitates the online formation and automatic cross-linking of microbeads for three-dimensional cell culture.[11] However, costly agents such as perfluorinated polyetherspolyethyleneglycol (PFPE-PEG) and 1H,1H,2H,2H-Perfluoro-1-octanol (PFO) were utilized for gelation and demulsification, respectively. Nonetheless, there have been very few studies that have developed a consumption-saving and high-throughput platform suitable for resource-constrained environments.

In this study, we present the production of homogeneous alginate hydrogel microbeads through a symmetric flow-focusing design implemented in a PDMS (polydimethylsiloxane) microfluidic device. This two-step methodology effectively addresses several limitations: (i)This method enables the precise and cost-effective generation of monodisperse alginate hydrogel droplets using a continuous phase of pure corn oil, thereby eliminating the need for surfactants; (ii) it reduces droplet coalescence in corn oil by incorporating only acetic acid, which facilitates the mass production of homogeneous alginate hydrogel microbeads; (iii) alginate hydrogel microspheres can be separated from corn oil in an oscillation state, rather than requiring a demulsifier phase.

Figure 1(a) illustrates the prototyping of the microfluidic device used for the formation of alginate microspheres. In our experiment, a calcium-EDTA complex dissolved in Na-alginate solution served as the dispersed phase, while corn oil (without surfactant) functioned as the continuous phase. These phases were infused into a PDMS microfluidic device, which was fabricated using a facile and low-cost method described in our previous study[12]. Droplets were generated at one of the flow-focusing microfluidic devices, where the cross-section of all microchannels was designed to be 350 × 350 μm². The distance of the double flow-focusing junction is designed to 10mm. The corn oil, devoid of surfactant, effectively produced droplets of the Na-alginate solution containing the calcium-EDTA complex, resulting in an emulsion with a narrow distribution in droplet size shown in Figure 1(b). At another flow-focusing junction downstream, acetic acid (2% w/w) dissolved in corn oil was introduced and mixed with the stream of monodisperse droplets. The $H^+$ ions from the acidic oil readily diffused into the aqueous Na-alginate droplets. Notably, the acidic oil was mixed with the corn oil downstream to stabilize the Na-alginate droplets against coalescence, thereby preserving the homogeneous size of the spherical alginate microspheres.

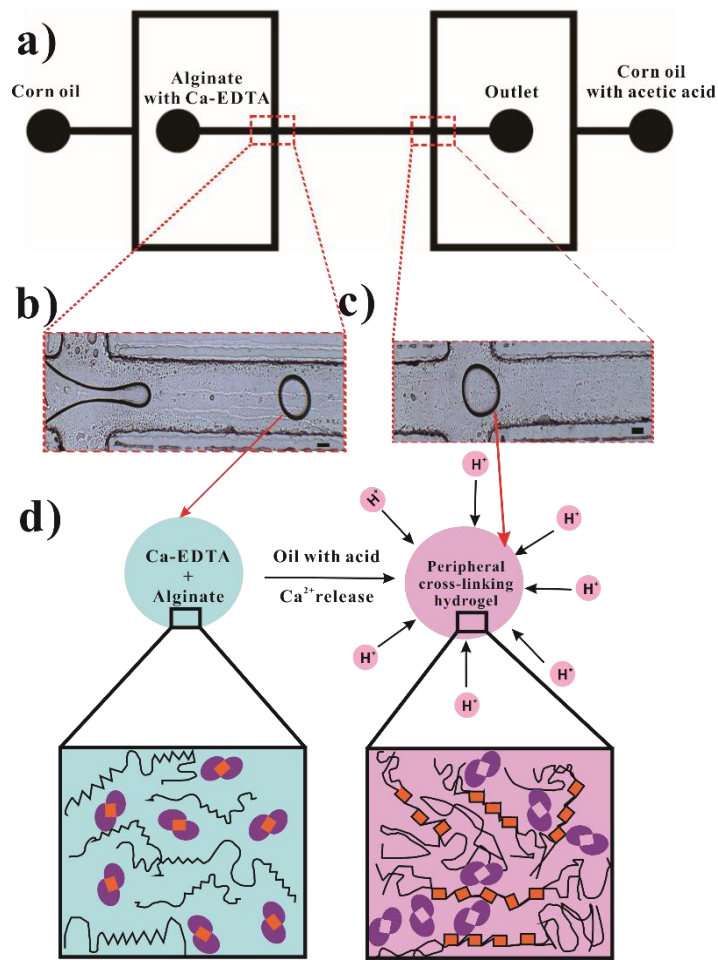

**Figure 1**. The microfluidic formation of homogeneously alginate microspheres involves several key components: a) A symmetric flow-focusing design of the microfluidic device facilitates the generation of alginate microspheres by releasing calcium ions from a calcium-EDTA complex on demand. b) A microscopic image illustrates the formation of alginate droplets using pure corn oil as the continuous phase (scale bar: 100 µm). c) The generation of cross-linked alginate microspheres is achieved by using acid corn oil in the flow-focusing device (scale bar: 100 µm). d) A schematic illustration depicts the progression of alginate crosslinking, wherein the introduction of acid at the flow-focusing structure induces the complete formation of cross-linked microspheres following the release of calcium ions from the calcium-EDTA complex solution.

To extract these hydrogel microspheres from the corn oil, we introduced the emulsions into a petri dish filled with aqueous solution and initially separated them from the oil phase while in an oscillating state on a micro-plate shaker, as illustrated in Figure 2. Figure 2(a) shows the hydrogel microspheres encapsulated in the oil phase formed by the microfluidic device. These entrapped microspheres can easily transition from the oil medium to the aqueous medium, thereby quenching the gelation reaction simultaneously, as depicted in Figure 2(b). The excess oil remains on top of the aqueous phase and can be aspirated after the microspheres have settled. The entire process was completed within 5 minutes and was cost-effective.

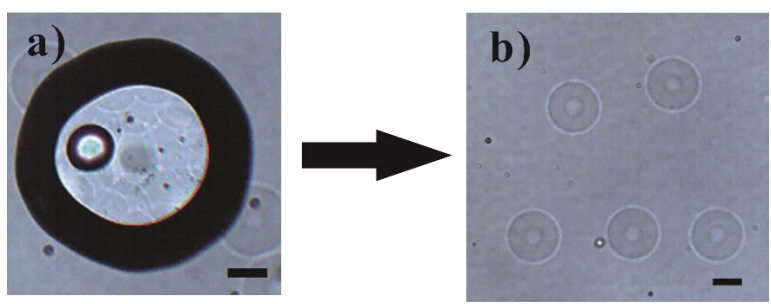

Figure 2. a) The microscopic image of alginate microspheres encapsulated in oil emulsion, b) The alginate microspheres were separated from oil droplet at the condition of oscillation state (scale bar: 100 µm).

Figure 3 illustrates that the dimensions of the hydrogel microspheres can be adjusted by varying the flow rate of the continuous phase, which is corn oil. At specific concentrations of the calcium-EDTA complex in an alginate medium (2 wt%) and corn oil with acetic acid (2% w/w), we obtained spherical hydrogel microspheres with diameters ranging from 135 μm to 182 μm across the varied continuous flow rates. In this step, the calcium-EDTA complex was mixed with an aqueous medium of calcium chloride (100 mM) and a solution of disodium-EDTA (100 mM) in equal ratios. The pH of this solution was adjusted to 7.2, ensuring that the complex remained highly stable. The dimensions of these microspheres exhibited an excellent coefficient of variation (C.V.) of less than 5%. We kept the flow rate of alginate solution constant (3 μL/min) while adjusting the flow rate of corn oil from 20 to 40 μL/min, thereby decreasing the emulsion diameter, as shown in Figure 3.

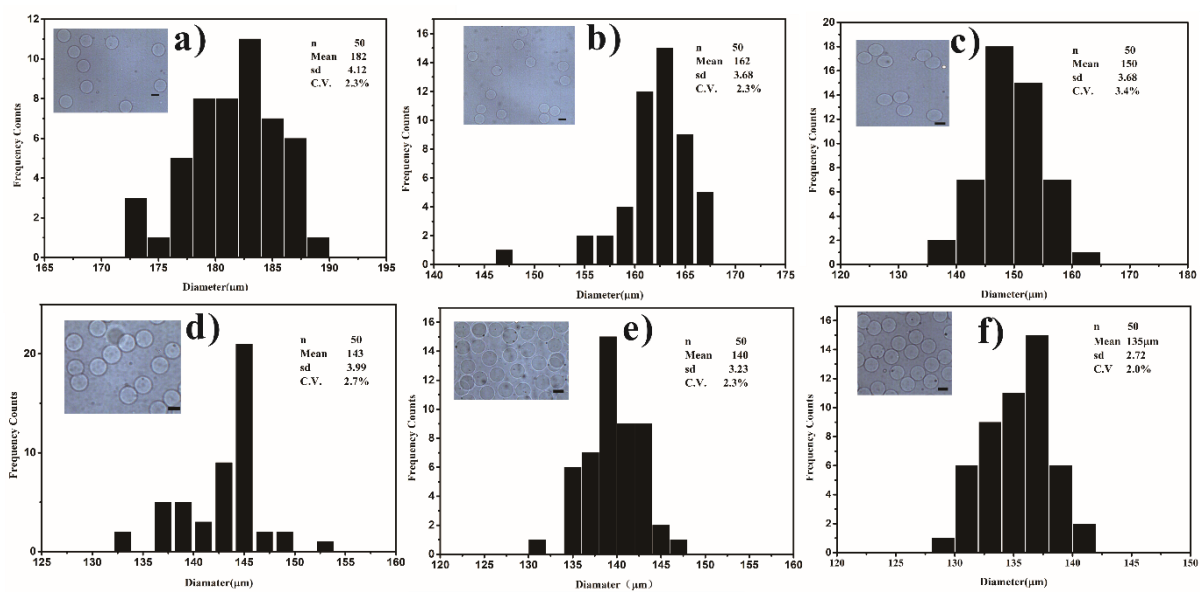

Figure 3. Bright-field microscopic images of alginate microspheres following demulsification. In these experiments, the flow rate of the disperse alginate–

calcium–EDTA solution is set at 3 µL/min. (a) The flow rate of the continuous oil phase is 20 µL/min to adjust the droplet size for the specific microfluidic device; (b) The flow rate of the continuous oil phase is 25 µL/min; (c) The flow rate of the continuous oil phase is 30 µL/min; (d) The flow rate of the continuous oil phase is 32 µL/min; (e) The flow rate of the continuous oil phase is 35 µL/min; (f) The flow rate of the continuous oil phase is 40 µL/min. All flow rates of the acidic oil are maintained at 40 µL/min (Scale bars: 100 µm).

In order to avoid the merging of alginate droplets without the use of surfactant, we incorporated corn oil mixed with acetic acid into the main flow after the droplets had formed. Previous studies have shown that the gelling interface is dynamic throughout the transition phase of the hydrogel. Additionally, the release of $Ca^{2+}$ and the gelation process are further complicated by a boundary that moves sequentially.[13] At room temperature, we produced spherical alginate microspheres with average diameters ranging from 135 µm to 182 µm, which varied according to the flow rates of the continuous oil phase. These microspheres demonstrated a high degree of uniformity in diameter, as evidenced by the coefficient of variation (C.V.) values presented in Figures 3a-f. In this step, a flow rate of corn oil with acetic acid set below 30 µL/min was insufficient for the successful generation of monodisperse droplets.

In summary, we present a novel two-step route for the microfluidic fabrication of monodisperse alginate hydrogel microspheres with ideal spherical homogeneity. The use of calcium-EDTA complex as a crosslinking precursor and $Ca^{2+}$ ions released by pH reduction resulting in the uniform gelation of the microspheres. This method will be enabled to be a precise tuning of the physical properties of the microgel through control of the amount of crosslinker and the nature of the alginate chains. We demonstrated that this procedure can be a cost-effectively and suitable for the resource-limited regions.

Research shown in this work was supported by The Scientific Research Foundation of Hunan Provincial Education, China (No. 23B0725 and 22A0549),The Foundation of Hunan Natural Science, China(2022JJ30467 and 2022JJ50312) the Research Foundation of Huaihua university, China (Grant HHUY2021-01, and HHUY2021-02), the Foundation of Hunan Double First-rate Discipline Construction Projects (Grant No. SWGC-04) and National Training Program project of Innovation and Entrepreneurship for Undergraduates (No. S202110548078).

**Conflicts of interest**

There are no conflicts to declare.

**Notes and references**